# Investigating transfer of learning in advanced quantum mechanics

Alexandru Maries[1], Ryan Sayer[2] and Chandralekha Singh[2]

[1]*University of Cincinnati, Department of Physics, Cincinnati, OH 45221*
[2]*University of Pittsburgh, Department of Physics and Astronomy, Pittsburgh, PA 15260*

**Abstract.** Research suggests that students often have difficulty transferring their learning from one context to another. We examine upper-level undergraduate and graduate students' facility with questions about the interference pattern in the double-slit experiment (DSE) with single photons and polarizers of various orientations placed in front of one or both slits. Before answering these types of questions, students had worked through a tutorial on the Mach-Zehnder Interferometer (MZI) in which they learned about interference of single photons when polarizers of various orientations are placed in the two paths of the MZI. After working on the MZI tutorial, students were asked similar questions in the isomorphic context of the DSE. We discuss the extent to which they were able to transfer what they learned in the context of the MZI to analogous problems in the isomorphic context of the DSE.

## I. INTRODUCTION

Transfer of learning from one context to another is a hallmark of expertise. Prior studies have often found that students have difficulty in transferring learning from one isomorphic problem to another which has a different context but involves identical physics principles [1]. For example, while angular momentum conservation implies that both a spinning skater and slowly spinning neutron star would speed up when their moment of inertia decreases, students may not discern the isomorphism and transfer their learning, e.g., from the skater problem to correctly answer the neutron star problem [2], even if the two problems are posed back to back. In quantum mechanics, additional difficulties can arise due to the unintuitive and abstract nature of the subject [3-7].

The goal of this study was to investigate the extent to which students are able to transfer learning from the context of the Mach-Zehnder Interferometer (MZI) with single photons [8] to the isomorphic context of the double-slit experiment (DSE) [9]. The two experiments selected for this study are useful because they can be used to illustrate fundamental principles of quantum mechanics and the underlying principles used to predict interference in both experiments is the same. To investigate transfer, we first designed several analogous pairs of questions in the MZI and DSE contexts by carefully examining the types of questions students answered when they worked on a research-based tutorial in the context of the MZI. After students worked on the MZI tutorial and answered the corresponding post-test questions, they were asked analogous questions in the DSE context (as part of a pre-test on the DSE) and we investigated the extent to which they were able to transfer their learning from the MZI context to the DSE context. We begin by providing a summary of the various aspects of the MZI and DSE experiments relevant for this study and discuss why these problems are isomorphic.

## II. ISOMORPHISM BETWEEN MZI AND DSE WITH SINGLE PHOTONS

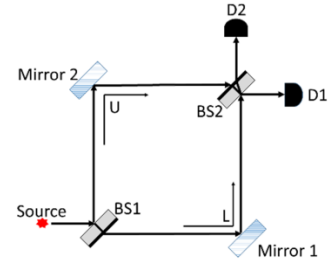

**FIG 1.** Basic MZI setup.

To understand the isomorphism between the MZI and DSE, we first consider the most basic MZI setup shown in Fig. 1 (BS1 and BS2 are beam splitters; BS1 is oriented such that it puts each single photon emitted from the source into an equal superposition of U and L path states shown, mirrors are for proper alignment and BS2 ensures that the components of the single photon state from both the U and L paths can be projected into each (photo) detector D1 and D2 after BS2 so that constructive, destructive, or intermediate interference can be observed at the two detectors D1 and D2 in Fig. 1 (depending on the path length difference between the U and L paths, e.g., if a phase shifter is placed in one path and its thickness is varied). If an *additional* detector is placed anywhere in the lower path L between BS1 and BS2 (not shown in Fig. 1), after encountering the detector, the superposition of the U and L path states of the photon collapses and if the photon does not get absorbed by that detector, the state of a photon inside of the MZI is the upper path state |U⟩. Conversely, if an additional detector is placed in the upper path U, after encountering that detector, if the photon is not absorbed by that detector, the state of a photon inside the MZI collapses to the lower path state $|L\rangle$. In these situations, when we have a detector in the U or L path of the MZI, if a photon is detected in the detector D1 or D2 after BS2, we have "which-path information" (WPI) about whether the photon took the U or L path of the MZI and no interference is observed at D1 or D2 [6,7,10]. If instead, no detector is placed in either path of the MZI (as in Fig. 1), the state of a photon in the MZI remains an equal superposition of the U and L path states after BS1, so we do not have WPI for the single photons and hence we do observe interference at the detectors D1 and D2.

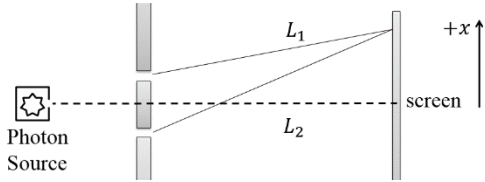

**FIG 2.** Basic DSE setup with single photons.

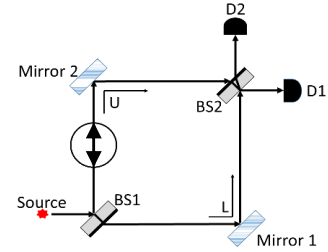

**FIG 3.** MZI setup with vertical polarizer in upper path

Now let us consider the DSE setup shown in Fig. 2. If slit 2 is blocked, the state of a photon inside the DSE (right after passing through the slits) is $|\Psi_1\rangle$ and if slit 1 is blocked, the state of the photon is $|\Psi_2\rangle$. If this photon is detected at the screen when one slit is blocked (the screen is the detection device in the DSE equivalent to photo-detectors D1 and D2 in the MZI), we have WPI about which slit the photon went through to arrive at the screen and hence no interference is observed. If neither slit is blocked, the state of a photon is an equal superposition of $|\Psi_1\rangle$ and $|\Psi_2\rangle$. In other words, the $|U\rangle$ and $|L\rangle$ path states in the MZI are analogous to the slit states $|\Psi_1\rangle$ and $|\Psi_2\rangle$ in the DSE. In the situations in which there is no detector in either path of the MZI and neither slit is blocked for the DSE, we do *not* have WPI and each photon interferes with itself at the screen and an interference pattern emerges after a large number of single photons are detected.

Now consider a situation in which the source emits 45° polarized single photons and we place a vertical polarizer in the upper path of the MZI (Fig. 3), which is analogous to placing a vertical polarizer in front of slit 1 in the DSE. We now have to use a four dimensional product space: a two dimensional space for path/slit states, $|U\rangle,|L\rangle/|\Psi_1\rangle,|\Psi_2\rangle$, and a two dimensional space for polarization states, for which a convenient basis is $\{|V\rangle, |H\rangle\}$ (vertical, horizontal polarization states). If a vertical polarizer is placed in the upper path of the MZI, the $|U\rangle$ state will be associated with a vertical polarization state ($|U\rangle|V\rangle$) and the $|L\rangle$ state is still associated with both vertical and horizontal polarization states ($|L\rangle|V\rangle + |L\rangle|H\rangle$). Thus, we have WPI for horizontally polarized photons detected at detectors D1 and D2, because the horizontal polarization is associated only with the $|L\rangle$ state, and we do not have WPI for the vertically polarized photons because the vertical polarization is associated with both the $|U\rangle$ and $|L\rangle$ states. Therefore, the photons that are detected in the $|V\rangle$ state interfere and those in the $|H\rangle$ state do not. In the DSE, the situation is analogous; if a vertical polarizer is placed in front of slit 1, horizontally polarized photons detected at the screen will not interfere, while vertically polarized photons will show interference.

While the contexts are isomorphic, the "surface" features of these two experiments are rather different. In the MZI, the paths are restricted and the photons arrive at point detectors D1 and D2, while in the DSE the photons are delocalized in the space between the slits and the screen and can be detected anywhere on the extended screen. In addition, in the DSE, there is no optical element corresponding to BS2 in the MZI (which mixes the components of the photon state from the two paths). These differences suggest that the surface features of these problems are quite different which can make it challenging for novices [1] to recognize the isomorphism. Thus, transfer of learning from one context to another is not guaranteed a priori even if students understand the underlying physics principles in the MZI context.

### III. METHODOLOGY

The participants in this study were 43 undergraduate students enrolled in upper-level quantum mechanics and 41 first year physics graduate students enrolled in a semester-long TA training course. For the undergraduate students, the MZI tutorial, MZI pre-/post-tests and the analogous DSE questions (given as part of a pre-test on the DSE) were graded for correctness. The graduate students worked on the tutorial and pre/post-test in the TA training course and while they also completed all of the materials, their work was only graded for completeness since the TA training course performance was graded as satisfactory or unsatisfactory.

The research-based MZI tutorial that students worked on before answering the transfer questions about the DSE uses thought experiments and an interactive simulation. The tutorial helps students learn topics such as the wave-particle duality of single photons, interference of a single photon with itself, probabilistic nature of quantum measurements, and collapse of a quantum state upon measurement. Students manipulate the MZI setup in the interactive simulation to predict and observe what happens at the detectors for various setups. They learn how placing polarizers of various orientations in the paths of the MZI affects the pattern observed at the detectors by considering whether a particular setup is such that WPI can be obtained for photons of certain polarizations that are detected at the detectors D1 and D2.

The pre-/post-test includes questions in which polarizers of various orientations are placed in one or both paths of the MZI. After students worked on the MZI tutorial along with the pre-/post-test (both graduate and undergraduate students performed very well on the post-test), they were given a pre-test on the DSE which included analogous questions in the DSE context, (referred to as the "DSE transfer questions"). The DSE transfer questions can be answered by using WPI reasoning that students learned in the context of the MZI. To investigate transfer of learning from the MZI to the DSE context, we compared student performance on the MZI pre-test questions with performance on the analogous DSE transfer questions by looking at frequency of correct responses and usage of WPI reasoning.

**DSE transfer questions**

The transfer questions for the DSE, which are analogous to situations in the MZI, are as follows: You perform a DSE in which photons that are polarized at +45° are sent one at a time towards the double slit. The wavelength of the photons is comparable to the slit width and the separation between the slits is more than twice the slit width. In all questions, assume that the same large number $N$ of photons reach the screen. In each situation, describe the pattern you expect to observe on the screen. Explain your reasoning.
(1) Situation described above. (2) Vertical polarizer placed in front of one slit. (3) Vertical polarizer placed in front of each slit. (4) Vertical and horizontal polarizer placed in front of slit 1 and 2, respectively. (5) A vertical and a horizontal polarizer placed in front of slit 1 and 2, respectively. Additionally, a polarizer making an angle of 45° with the horizontal is placed in between the slits and the screen.

These questions are analogous to those asked in the context of the MZI, e.g., situation 2 above is analogous to the MZI setup shown in Fig. 3. It is important to keep in mind that the DSE transfer questions above were part of a larger pre-test about the DSE which had other questions related to the DSE with single particles, and the MZI pre-/post-tests also involved many other questions in other situations (e.g., effect of removing BS2 on interference at the detectors D1 or D2) and included other types of questions (e.g., the percentage of photons of a given polarization arriving at D1 and D2) not included in the transfer questions.

We hypothesize that the difference between how students perform on these questions in the MZI pre-test and in the DSE transfer context may provide an indication of the extent to which they transfer their learning from the MZI tutorial to the analogous situations in the DSE. We note that the first question is an exception because students may know from prior courses that in the basic DSE setup, an interference pattern is observed when no polarizers are used.

## IV. RESULTS

Table I shows undergraduate and graduate students' average performance (as measured by frequency of correct answers) on the MZI pre-test and DSE transfer questions described above. Situations 1 and 3 were not asked in the same form in the context of the MZI (i.e., these questions asked about the fraction of photons that would be detected by D1 and D2 which is somewhat different although closely related to the presence/absence of interference) so responses to those in the MZI context are not included. The $p$ values for comparison between the analogous questions in the two contexts in Table I show that on average, on the DSE transfer questions, students performed significantly better than on the analogous MZI pre-test questions, suggesting positive transfer of learning between the two contexts.

**TABLE I.** Average performance (out of a maximum of 1) of undergraduates (US) and graduate students (GS) on questions related to the effect of polarizers on the interference pattern in MZI and DSE contexts, and $p$ value for comparison of the performance in the two contexts.

| Question # | | 1 | 2 | 3 | 4 | 5 |
|---|---|---|---|---|---|---|
| US | MZI | | 0.16 | | 0.27 | 0.20 |
|  | DSE | 0.91 | 0.50 | 0.71 | 0.81 | 0.67 |
|  | $p$ val | | 0.004 | | <0.001 | <0.001 |
| GS | MZI | | 0.24 | | 0.42 | 0.41 |
|  | DSE | 0.97 | 0.48 | 0.81 | 0.84 | 0.77 |
|  | $p$ val | | 0.029 | | <0.001 | 0.001 |

Another indication of transfer is that students often used WPI reasoning learned in the MZI tutorial to answer the DSE transfer questions. Table II shows the percentage of both undergraduate and graduate students who used reasoning related to WPI on the DSE transfer questions 2 to 5 among those who provided reasoning for their answers. On these questions, a majority of the students who used reasoning related to WPI used it correctly. In contrast, nearly no student used such reasoning on the MZI pre-test (these percentages are almost zero and not included in Table II).

For example, on DSE transfer question 2, one student wrote: "The interference pattern will be fuzzier because we have which path data for any photons that are not vertically polarized" (this is a typical response of students who used WPI reasoning) and another wrote "Interference for the horizontally polarized components. But no interference for the vertically polarized as we have which-path [info]. We know which slit the vertically polarized photons originated from." The first student appropriately transferred his learning to the DSE context, while the second is confused about the polarization for which WPI is known. However, he did attempt to transfer reasoning related to WPI learned in the MZI to the DSE context. Including both undergraduate and graduate students, 58% of those who used WPI related reasoning answered question 2 correctly (8 students). We also note that this question was the most challenging question for both undergraduate and graduate students.

On question 3, the most common student reasoning related to WPI is that only vertically polarized photons are detected at the screen and there is no WPI for those photons. This or very similar correct reasoning was used by 75% of students who used WPI to explain their reasoning. Students rarely used WPI reasoning incorrectly on this question.

On question 4, all students who used WPI reasoning answered the question correctly, indicating that all students who attempted to transfer learning from the MZI tutorial to the DSE context did so appropriately (i.e., reasoned that there is WPI for both vertically and horizontally polarized photons detected at the screen, which leads to no interference).

**TABLE II.** Percentage of undergraduates (US) and graduate students (GS) who used reasoning related to WPI out of those with reasoning on DSE transfer questions 2-5.

| Question # | 2 | 3 | 4 | 5 |
|---|---|---|---|---|
| US | 37 | 37 | 57 | 62 |
| GS | 33 | 20 | 60 | 44 |

The situation on question 5 in the DSE (vertical polarizer in front of one slit, horizontal polarizer in front of the other, +45° polarizer in front of the screen) is a "quantum eraser" [10] because the +45° polarizer erases WPI that could be obtained from the other two polarizers alone. Therefore, an interference pattern forms on the screen. The analogous situation is discussed in the MZI tutorial and students learned the term "quantum eraser" in that context. When answering this question in the DSE transfer context, the majority of students used reasoning related to WPI and 82% answered the question correctly showing appropriate transfer. In addition, 25% of the undergraduates and 10% of the graduate students mentioned "quantum eraser", which was a term learned in MZI context, or drew the parallel between the DSE and MZI (e.g., answers to DSE questions included comments such as "this is just like the MZI situation in which…").

**Possible reasons for observed transfer**

While it is difficult to identify the exact causes of the substantial transfer from the MZI to the DSE context, we hypothesize that the following may play a role:

1) Upper-level undergraduate and graduate students have developed sufficient abstract reasoning skills which allows them to recognize the isomorphism between these situations and the usefulness of reasoning about WPI in both contexts.

2) While the isomorphism between the MZI and DSE is in underlying physics and the contexts are different, both use single photons and polarizers of various orientations placed in front of one or both slits or paths. This type of similarity may have prompted students to utilize analogous reasoning when answering the DSE transfer questions. We note, however, that in the MZI post-test, student average scores were near the ceiling (~ 90%), while the averages on the DSE transfer questions were around 70% for both undergraduates and graduate students, implying that the transfer from the MZI to the DSE context is not perfect. In addition, the questions on both the MZI and DSE discussed here were part of longer pre-/post-test/quiz on these experiments which asked about other situations and included other types of questions. For example, in the DSE quiz, students answered 13 additional questions along with the transfer questions.

3) After students submitted the completed MZI tutorial and took the related post-test, the DSE transfer questions were given in the following class. This proximity in timing may make it more likely for students to be able to discern the similarity between the two contexts and transfer their learning from the MZI context to the DSE context. However, as mentioned earlier, in introductory physics, even if two questions which require use of the same underlying physics principles are asked back to back as part of the same quiz, a majority of students may not discern the similarity between the questions and answer them using different reasoning [2].

## V. SUMMARY

In this study, we find evidence that advanced students can transfer learning from a tutorial on the MZI to an isomorphic context in the DSE without an explicit intervention to aid them in this regard. The MZI tutorial introduced students to the concept of WPI and guided them to use it to reason about whether or not interference is observed at the detectors in a particular MZI setup. When the DSE transfer questions were administered, students performed significantly better on the DSE transfer questions (average above 70%) than on the analogous MZI pre-test questions (average ~35%). Another indication of transfer is that students often explicitly used reasoning learned in the context of the MZI to answer the DSE transfer questions, e.g., they used reasoning related to WPI, and most students who used this type of reasoning did so correctly, indicating appropriate transfer from the MZI to the DSE context. In addition, students sometimes drew the parallel between the DSE and the MZI themselves.

Possible reasons for the observed transfer include the close temporal proximity of the MZI tutorial to the DSE transfer questions and the fact that both the MZI and DSE questions relate to single photons and polarizers in front of various paths/slits. However, as noted earlier, introductory students often have difficulty discerning the similarity between isomorphic problems even if they are placed back to back [2]. In addition, the differences between the setups suggest that the surface features of these problems are quite different which can make it challenging to recognize the isomorphism [1]. Therefore, it is encouraging that advanced students have developed sufficient reasoning skills to be able to transfer their learning at least in the context discussed.

## VI. ACKNOWLEDGEMENTS

We thank the National Science Foundation for awards NSF-PHY-1202909 and NSF-PHY-1505460.